\documentclass[12pt,preprint]{aastex}

\usepackage{url}
\usepackage{amsmath}
\usepackage{color}
\usepackage{textcomp}
\usepackage{amssymb}
\usepackage{wasysym}

\newcommand{\ssst}{\scriptscriptstyle}
\newcommand{\HI}{{\ion{H}{1}}}
\newcommand{\ps}{\,{\rm s}^{-1}}
\newcommand{\km}{\,{\rm km}}
\newcommand{\cm}{\,{\rm cm}}
 
\newcommand{\gray}{{\rm $\gamma$-ray}}
\newcommand{\fermi}{\textit{Fermi}}
\newcommand{\MHz}{\,{\rm MHz}} 
\newcommand{\twCO}{$^{12}{\rm CO}$} 
\newcommand{\Jotz}{$J$=1--0}  
\newcommand{\Jyperb}{\,{\rm Jy}\,{\rm beam}^{-1}}
\newcommand{\VLSR}{V_{\ssst\rm LSR}}
\newcommand{\du}{d_{12}}

\newcommand{\nt}{n_{\rm t}}

\begin{document}

\title{GeV \gray\ Emission Detected by \fermi-LAT
Probably Associated with the Thermal Composite Supernova Remnant Kesteven~41 in a Molecular Environment}

\author{Bing Liu \altaffilmark{1},
        Yang Chen \altaffilmark{1,2,5},
        Xiao Zhang \altaffilmark{1},
        Gao-Yuan Zhang \altaffilmark{1},
        Yi Xing \altaffilmark{3},
     \& Thomas G. Pannuti\altaffilmark{4}}

\altaffiltext{1}{\footnotesize Department of Astronomy, Nanjing University,
163 Xianlin Avenue, Nanjing 210023, China}

\altaffiltext{2}{\footnotesize 
Key Laboratory of Modern Astronomy and Astrophysics,
Nanjing University, Ministry of Education, Nanjing 210093, China}

\altaffiltext{3}{\footnotesize 
Key Laboratory for Research in Galaxies and Cosmology,
Shanghai Astronomical Observatory, Chinese Academy of Sciences,
80 Nandan Road, Shanghai 200030, China}

\altaffiltext{4}{\footnotesize Space Science Center, 
Department of Earth and Space Sciences, 
Morehead State University, 235 Martindale Drive, Morehead, KY 40351, USA}

\altaffiltext{5}{\footnotesize Corresponding author.}

\begin{abstract}
Hadronic emission from supernova remnant (SNR)--molecular cloud
(MC) association systems has been widely regarded as a probe
of the shock-accelerated cosmic-ray protons.
Here, we report on the detection of a $\gamma$-ray emission
source with a significance of $24\sigma$ in 0.2--300\,GeV,
projected to lie to the northwest of the thermal composite SNR~Kesteven~41,
using 5.6 years of \fermi-Large Area Telescope (LAT) observation data.
No significant long-term variability in the energy range 0.2--300\,GeV
is detected around this source.
The $3\sigma$ error circle, $0\fdg09$ in radius,
covers the 1720MHz OH maser
and is essentially consistent with
the location of the $\VLSR\sim-50\km\ps$ MC with which
the SNR interacts.
The source emission has an exponential cutoff power-law spectrum  
with a photon index of $1.9\pm0.1$ and a cutoff energy of $4.0\pm0.9\ {\rm GeV}$,
and the corresponding 0.2--300\,GeV luminosity is
$\sim1.3\times 10^{36}\,\mbox{erg}\ps$ at a distance of 12\,kpc.
There is no radio pulsar in the $3\sigma$ circle
responsible for the high $\gamma$-ray luminosity.
While the inverse Compton scattering scenario would
lead to a difficulty in the electron energy budget,
the source emission can naturally be explained by the
hadronic interaction between the relativistic protons
accelerated by the shock of SNR~Kesteven~41 and the
adjacent northwestern MC.
In this paper, we present a list of Galactic thermal composite SNRs detected
at GeV \gray\ energies by \fermi-LAT is presented.

\end{abstract}


\section{INTRODUCTION}
\label{sec:intro}

The origin of the cosmic rays (CRs) remains one of the most debated
issues in high energy astrophysics more than 100 years after
they were detected for the first time.  
Supernova remnants (SNRs), whose strong shocks contain huge amounts of energy,
are considered to be the most probable candidates among the Galactic CR acceleration
sources (e.g., \citealt{Ginzburg1969ocr}). 
A long standing argument concerning the putative SNR-CR link is that 
non-thermal radio emission from SNRs provides clear evidence for electron
acceleration, whereas the CR spectrum observed on Earth is 99$\%$ protons
and other nuclei.
Therefore, the \gray\ emission from SNRs that is dominated by the decay of $\pi^0$
mesons produced via proton-proton collisions (i.e., the hadronic interaction)
plays a key role in providing
evidence for proton acceleration \citep{Ackermann2013Sci}.
However, it is often difficult to distinguish between the hadronic
\gray\ emission and the electrons' inverse Compton or non-thermal 
bremstrahlung emission (i.e., the leptonic emission).
There are generally two scenarios that describe how hadronic {\gray}s
are produced in SNRs. 
In one scenario, the $\pi^0$-decay emission is suggested to arise
from shock-crushed dense clouds where the accelerated protons frozen
in the clouds efficiently collide with target cloud gas
\citep[e.g.,][]{Blandford1982,Uchiyama2010, TangChe2014}. 
In the other scenario, the hadronic {\gray}s are ascribed to 
interactions between the relativistic protons escaping from the SNR
shock and adjacent molecular clouds (MCs; \citealt[e.g.,][]{Aharonian1996AA,
Gabici2009, lichen2010, Ohira2011}). 
In both scenarios, the SNRs interacting with MCs are 
crucial probes in the search for the signatures of proton acceleration.
The hadronic \gray\ emission from SNR-MC systems is usually bright
around GeV, and a series of GeV-bright SNRs interacting with MCs
have recently been discovered with the Large Area Telescope (LAT) 
on board the \textit{Fermi Gamma-ray Space Telescope}. 
These SNRs include W51C \citep{Abdo2009W51C}, W44 \citep{Abdo2010W44SCi}, 
IC~443 \citep{Abdo2010IC443}, W28 \citep{Abdo2010W28},
W41 \citep{Castro2013}, RCW~103 \citep{XingRCW103}, etc.
Additional GeV observations continue to enlarge the sample of hadronic
interaction between SNRs and MCs, 
and here we present a GeV study of another SNR, namely, Kesteven~41(G337.8$-$0.1).

As a southern-sky SNR, Kes~41 
is shown to be centrally brightened in X-rays within a distorted radio shell 
by a \textit{XMM}-\textit{Newton} observation \citep{Combi2008}, 
and therefore is classified as a thermal composite
(or mixed-morphology) SNR \citep{Jones98,Rho98}.
The X-ray emitting plasma of the SNR has been newly revealed to be rich in sulfur and argon;
thus, Kes~41 joins the subclass of ``enhanced-abundance" or ``ejecta-dominated"
thermal composites \citep{gaoyuan2015}. 
Kes~41 has also been found to be interacting with an adjacent MC,
as indicated  by the 1720$\MHz$ hydroxyl radical (OH) maser emission 
detected in the northern radio shell \citep{Koralesky1998,Caswell2004}.
Recently, we found that Kes~41 is associated with a giant MC
at a systemic local standard of rest (LSR) velocity of $-50\km\ps$ 
and is confined in a cavity delineated by a northern molecular shell, 
a western concave MC, and a southeastern \HI\ cloud \citep{gaoyuan2015}.
The forward shock is suggested to have left the adiabatic stage
since the SNR shock encountered the cavity wall,
while the inner thermal X-rays are ascribed to heating
by the reflection shock from the cavity wall.
The birth of Kes~41 inside the molecular cavity provides a mass
estimate of $\ga18M_{\odot}$ for the stellar progenitor.
It is logical and meaningful to search for the hadronic emission
due to the interaction of the SNR with the dense environmental gas.

In this paper, we report the results from a spatial and spectral
analysis of the \fermi-LAT observation data of the Kes~41 region.
We describe the \fermi\ observation data in Section \ref{sec:obs},
and present the data analysis and results in Section \ref{subsec:ar}.
The possible physical relation of the detected \gray\ emission
with the SNR is discussed in Section \ref{sec:dis}.

\section{OBSERVATIONS AND DATA REDUCTION}

\label{sec:obs}

The LAT on board \fermi, launched on 2008 June 11,
is a $\gamma$-ray imaging instrument 
that covers a very wide range of energy from 20 MeV and up to 300 GeV.
It reconstructs the direction of incident $\gamma$-rays
by tracking the electrons and positrons resulting from pair conversion 
of the $\gamma$-rays in the solid state silicon trackers,
and measures the energy of the subsequent electromagnetic showers 
that develop in the cesium iodide calorimeters 
\citep{Atwood2009fermilat}.
The point-spread function (PSF) varies largely with photon energy 
and improves at high energies (the 68\% containment radius at $>$2 GeV 
is smaller than 0\fdg5 \citealt{Atwood2009fermilat}).  

We use the reconstructed Pass 7 reprocessed version 
of 5.6 years of accumulated \fermi-LAT data\footnote{http://fermi.gsfc.nasa.gov/ssc/data} 
that has been selected from 2008 August 04 15:43:37 (UTC) to 2014 April 01 02:29:28 (UTC).  
We analyze the data with the standard software, 
\emph{ScienceTools} version {\tt v9r32p5}\footnote{See http://fermi.gsfc.nasa.gov/ssc} 
released on 2013 October 24, 
with the instrument response functions (IRFs) P7REP\_SOURCE\_V15. 
Standard selection criteria are applied to the data selection process
as described below. 
The \emph{Source} (evclass$=$2) events are selected 
and the maximum zenith angle cut is 100$^{\circ}$ 
to reduce the residual $\gamma$-rays from CR interactions in the upper atmosphere.
We used the standard criteria for selecting time intervals for analysis:
(DATA\_QUAL==1) \&\& (LAT\_CONFIG==1) \&\& ABS(ROCK\_ANGLE)$<$52.
The analysis is restricted to the energy range above 200 MeV due to 
uncertainties in the effective area and broad PSF at low energies
and below 300 GeV due to limited statistics.

\section{ANALYSIS AND RESULTS}
\label{subsec:ar}
In our analysis, we select the LAT events inside a $14\arcdeg\times 14\arcdeg$ 
region of interest (ROI, in equatorial coordinate system) 
centered at the position of the Kes~41 
(R.A. (J2000) = 16$^{\rm h}$39$^{\rm m}$00$^{\rm s}$ and decl. (J2000) = $-46^{\circ}58'59''$) 
with a bin size of 0$^{\circ}\!$.04 $\times$ 0$^{\circ}\!$.04.
We perform our analysis following the standard binned likelihood analysis procedure. 
The second \fermi-LAT Catalog (2FGL) sources \citep{Nolan20122FGL} 
within radius 15$^\circ$ around Kes~41 are included in the source model, 
which was generated by the user-contributed software {\tt make2FGLxml.py}\footnote{http://fermi.gsfc.nasa.gov/ssc/data/analysis/user/}.  
The Galactic and extragalactic diffuse background components 
(as specified in the files \emph{gll\_iem\_v05.fits} and
\emph{iso\_source05.txt, respectively}) are used.
In the likelihood fittings, the spectral parameters of the sources
located beyond 10$^\circ$ of the ROI center
are fixed to the values reported in 2FGL, 
and the spectral parameters of all the sources located 
within 10$^\circ$ of  the center of ROI, together with 
the normalizations of the two diffuse backgrounds, are allowed to vary.
The fittings are performed with the optimizer NEWMINUIT until
convergence is achieved. 

\subsection{Source Detection}
\label{subsec:si}

\begin{figure}
\centering
\includegraphics[width=0.85\textwidth]{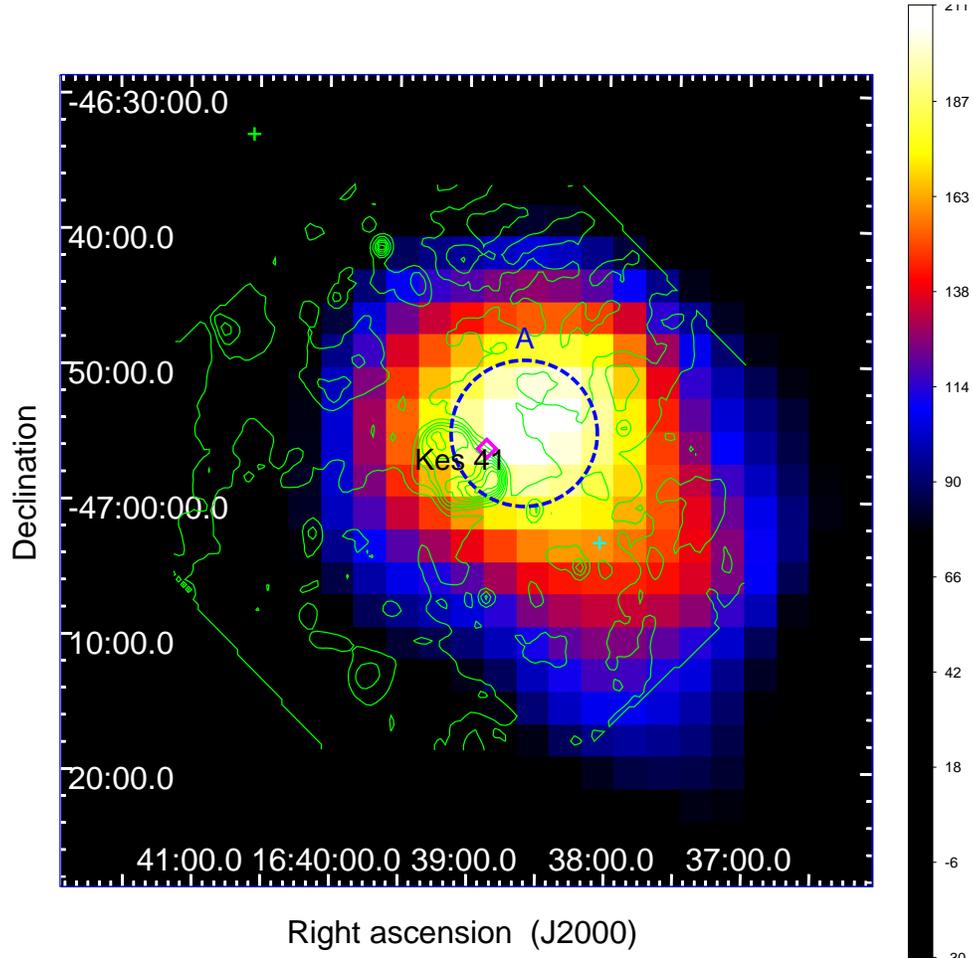}
\caption{TS map (2--300 GeV) of $1^{\circ}\times1^{\circ}$ region 
centered at Kes~41.
All sources except 2FGL~J1638.0$-$4703c have been subtracted. 
The green cross labels the position of a 2FGL source, 
the cyan cross labels the position of 2FGL~J1638.0$-$4703c, 
and the dashed blue circle indicates the 3$\sigma$ error range
of the best-fit position 
for the residual emission found in the Kes~41 region. 
The image is overlaid with the MOST 843$\MHz$ radio contours (in green)
(at seven linear scale levels between 0.00 and 0.79$\Jyperb$;
from \cite{Whiteoak1996}).
The magenta diamond represents the OH ($1720\MHz$) maser spot 
\citep{Koralesky1998}.}
\label{fig:tsmap}
\end{figure}

First, a binned likelihood analysis is applied in the energy range 
2--300\,GeV.
In the source model, the source 2FGL~J1638.0$-$4703c,
which is very close to Kes~41,
has been removed due to the uncertainty of
its spatial and spectral information
caused by the imperfectly modeled diffuse emission,
and thus needs to be treated with great care\footnote{http://fermi.gsfc.nasa.gov/ssc/data/access/lat/2yr\_catalog/}
\citep{Nolan20122FGL}.
A newly discovered \gray\ source (HESS~J1641$-$463, \citealt{Lemoine2014})
has been added assuming a power-law spectrum.
Then, the test statistic
(TS, defined as $2(\log{\mathcal L}-\log{\mathcal L}_0)$,
here ${\mathcal L}_0$ is the likelihood of null hypothesis and  ${\mathcal L}$ 
is the likelihood with the source included) map for
a $1^{\circ}\times1^{\circ}$ region 
centered at Kes~41 is made after subtracting this baseline model
(see Fig.~\ref{fig:tsmap}). 
As can be seen in Figure~\ref{fig:tsmap}, 
there is excess $\gamma$-ray emission in the region of Kes~41.
The position of the peak of the TS value 
is on the northwest of the SNR, but
does not agree with the position of 2FGL~J1638.0$-$4703c.
It is generally consistent with the location of the dense MC
at $\VLSR\sim-50\km\ps$ which is found to be associated with the SNR
\citep{gaoyuan2015}.
We also perform an analysis in the low energy range 0.2--2\,GeV, 
and some residual $\gamma$-ray emission is detected at the same position. 
Therefore, we add a point source with a power-law spectrum at the position 
where the TS value is highest in our source model
to approximate the excess emission. 
After that, we conduct a binned likelihood analysis
in the broad energy range 0.2--300\,GeV
and, utilizing \textit{gtfindsrc}
(a tool in the LAT software package \emph{ScienceTools}),
we find the best-fit position of the excess $\gamma$-ray emission
at (R.A. (J2000) = 16$^h$ 38$^m$ 36$^s\!$.00, 
decl. (J2000) = $-$46$^{\circ}$ 55$'$ 06$''\!$.96)
with $1\sigma$ nominal uncertainty of 0\fdg03
and $3\sigma$ nominal uncertainty of 0\fdg09. 

By comparison,
we detect this source as a point-like source with a power-law spectrum 
in 0.2--300 GeV with a significance of $24\sigma$ 
at the  best-fit position and increases the significance by $1\sigma$ 
over the position of 2FGL~J1638.0$-$4703c.
The data we use are collected from 5.6 years of \fermi-LAT observations
while the tentative source 2FGL~J1638.0$-$4703c was suggested
based on the first two years of observations.
Both the statistical result and increased exposure time hence suggest that
the $\gamma$-ray emission excess at the best-fit position adjacent to Kes~41
is more significant than 2FGL~J1638.0$-$4703c. 
Thus, we replace 2FGL~J1638.0$-$4703c with this new source
at the best-fit position 
(hereafter source A) in the following analysis.

In an attempt to explore the origin of the $\gamma$-ray emission of source A, 
we searched in the SIMBAD Astronomical Database \citep{simbad}
within a 3$\sigma$ error circle of the source (see Fig.~\ref{fig:tsmap}).
In addition to SNR Kes~41, only nine dark clouds, a young stellar object
candidate, and an infrared source are known to exist in the region. 
Therefore, the origin of this $\gamma$-ray emission is most likely related
to the SNR.

\subsection{Timing Analysis}
\label{subsec:tim}

\begin{figure}
\centering
\includegraphics[width=0.85\textwidth]{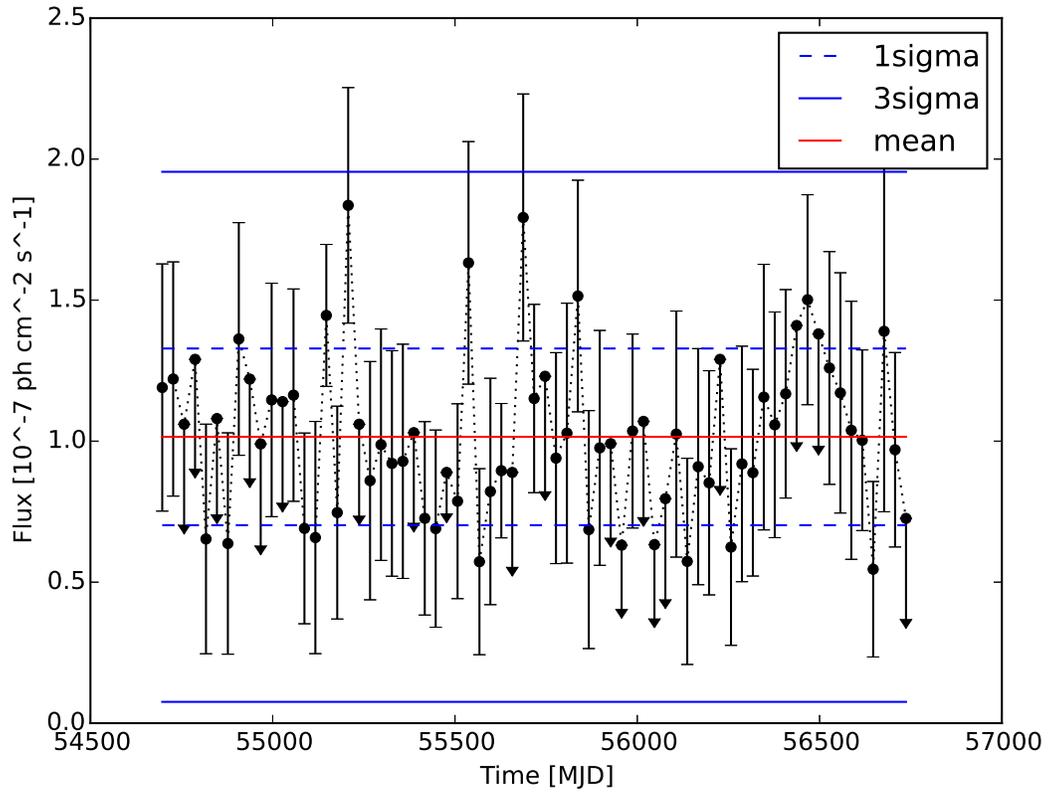}
\caption{Monthly \gray\ light curve of source~A 
in the energy range of 0.2$-$300 GeV.}
\label{fig:lc}
\end{figure}

We next search for long-term variability in the one month binned light curve 
of source~A in the energy range 0.2--300 GeV, which is obtained from 
likelihood analysis \citep{Nolan20122FGL} in each time bin.
As can be seen in the light curve (Figure~\ref{fig:lc}), 
all of the flux points remain within 3$\sigma$ uncertainties of the average flux.
Fitting the flux points with TS value $>4$  
to a constant flux model (shown as a red line in Figure \ref{fig:lc})
yields a $\chi^2$ $\sim$ 32.3 with 48 degrees of freedom (dof). 
Moreover, we calculate the Variability$\_$Index,
${\rm TS}_{\rm var}$, of source A (with all 69 time bins) 
in the 0.2--300\,GeV energy range
according to the method introduced in \S3.6 of \citet{Nolan20122FGL}.
If the flux is constant,
then ${\rm TS}_{\rm var}$ is distributed as $\chi^2$ with 68 dof, and
variability would be considered probable when ${\rm TS}_{\rm var}$
could exceed the threshold of 98.0 corresponding to 99$\%$ confidence.
The computed ${\rm TS}_{\rm var}$ of source A is 65.2,
corresponding to a confidence level $<50\%$ for a variable source.
These results suggest that there is no signifiant long-term variability 
observed in the region of Source A in the 0.2--300\,GeV energy range.
On the other hand, we construct 1000 s binned light curves of source~A in the same energy range
which are obtained through \fermi-LAT aperture photometry 
analyses\footnote{http://fermi.gsfc.nasa.gov/ssc/data/analysis/scitools/aperture\_photometry.html}
using LAT photons within different aperture radius from 0\fdg2 to 0\fdg5 .
We analyze these light curves for periodic signals, but no significant periodicity is detected.
However, this method is statistically limited and the periodicity is hard to detect due to the massive 
diffuse background photons in a low galactic latitude.

There is a close positional correspondence 
between source A, suggested here as a steady source,
and 3FGL J1638.6$-$4654, which is indicated as a variable source 
in the third \fermi-LAT Catalog (3FGL) \citep{Acero2015}.
3FGL J1638.6$-$4654 is detected in 0.1--300\,GeV with a significance
of $13\sigma$ and a 0.1--300\,GeV energy flux 
$\sim 7.3\times10^{-11}$ erg$\cm^{-2}\ps$;
and source~A has a higher significance ($24\sigma$)
with an energy flux of $\sim 7.5\times10^{-11}\,\mbox{erg}\cm^{-2}\ps$
in the 0.2--300\,GeV energy range (see Section \ref{subsec:pulsar};
the flux will be somewhat higher in 0.1--300\,GeV).
The use of  different spectral models and different energy ranges 
in the two timing analyses may contribute to the discrepancy in the variability 
between source A and 3FGL J1638.6$-$4654. 
The spectrum of source A is fit to a power-law model
and the spectrum of 3FGL J1638.6$-$4654 is fit to a log-parabola model.
Moreover, our timing analysis of source A uses photons in the energy range 0.2--300\,GeV
while 3FGL J1638.6$-$4654 is analyzed in the energy range 0.1--300\,GeV.
Our timing analysis would not be sensitive to flux variations 
(if any) below 0.2 GeV.

\subsection{Spatial Distribution Analysis}
\label{subsec:sda}

We analyze the spatial distribution of source~A, which is very likely
to be associated with Kes~41,
to examine whether it is a point-like or extended source. 
We apply both point-source and uniform-disk models with power-law spectra 
at the best-fit position to fit the emission in the energe range 2$-$300 GeV. 
In the point-source case, we set the spectral normalizations of the sources 
within 10$^{\circ}$ of Kes~41 as free parameters, 
and fix all the other parameters at the 2FGL values. 
A TS value of 207 is obtained.
In the disk case, the observed radius range for the uniform disks 
is 0\fdg1--0\fdg5 with a step of 0\fdg1. 
We fix all of the spectral parameters of the sources 
at the values obtained above, 
but allow the spectral normalization parameters of the disk models to be free parameters.  
The TS$_{ext}$ value (calculated from
$2\log({\mathcal L}_{\rm disk}/{\mathcal L}_{\rm point})$)
for each radius is smaller than zero,
while the extended source detection threshold is TS$_{ext} = 16$
\citep{Lande2012exted}, 
which implies that no significant extended emission is detected.
As a result, the GeV \gray\ emission from source~A seems to be point-like.

\subsection{Spectral Analysis}
\label{subsec:sa}

The $\gamma$-ray spectrum of source~A is extracted via
the maximum likelihood analysis of the LAT data 
in 6 divided energy bands from 0.2--300 GeV (see Table~\ref{tab:flux}). 
The spectral normalization parameters of the sources within $5^{\circ}$
of  Kes~41 are allowed to vary,
but all of the other source parameters are fixed.
In addition to the statistical uncertainties associated with 
the likelihood fits to the data, 
the uncertainty of the Galactic diffuse background intensity is considered. 
We vary the normalization of the Galactic background by $\pm6\%$ 
from the best-fit values at each energy bin
and estimate the flux from the object of interest 
using these new artificially frozen values of the background,
following the treatment in \citet{Abdo2009W51C}.
The possible systematic errors are estimated
to be 46\% (0.2--0.5 GeV), 40\% (0.5--1.0 GeV),
20\% (1.0--3.0 GeV), and $< 15\%$ ($>3$ GeV). 
We keep only spectral flux points with TS higher than 4 
(which corresponds to the detection significance of 2$\sigma$)
and derive 95\% flux upper limits in the other energy bins. 
The obtained spectral data for source~A are provided 
in Table~\ref{tab:flux}. 

We fit the 0.2--300~GeV spectral data of source~A
with a power-law model.
The obtained spectral shape is relatively flat
with a photon index of $\Gamma=2.38\pm0.03$.
The flux is $(9.2\pm1.0)\times10^{-11}
\,\mbox{erg}\cm^{-2}\ps$, corresponding to a luminosity
of $\sim1.6\times 10^{36}\du^2$ erg s$^{-1}$, 
where $\du=d/12$kpc is the distance to the MC associated with SNR Kes~41 
in units of the referecnce value estimated 
from the maser observation \citep{Koralesky1998}.
Also see Section \ref{subsec:pulsar} for an estimate of the flux and
luminosity with an exponential cutoff.

\section{DISCUSSION ON THE NATURE OF SOURCE A}
\label{sec:dis}

Based on our analysis of 5.6 years of \fermi-LAT data 
for the environment surrounding Kes 41, we have found a $\gamma$-ray source detected at 
a significance of $\sim 24\sigma$ that appears to 
be coincident with the northwest rim of Kes 41.

The relation between source A and Kes~41 is crucial 
for determining the origin of the \gray\ emission. 
In this section, we will discuss the possiblity of the \gray\
emission arising from a pulsar and an SNR-MC hadronic interaction,
respectively. 

\subsection{A Pulsar?}
\label{subsec:pulsar}
Galactic pulsars are important $\gamma$-ray source candidates,
and there have been numerous pulsars detected by \fermi-LAT
in recent years \citep{Abdo2010Pul}.
Although the $3\sigma$ error circle here does not include any known pulsars,
the possibility of correspondence to a pulsar associated with Kes~41 
still cannot be ignored.
Theoretically, there may be a descendent stellar compact remnant
after the core-collapse supernova (SN) explosion
of the $\ga18M_{\odot}$ progenitor of the remnant \citep{gaoyuan2015}.
Such a compact stellar remnant has not
been conclusively associated with Kes 41 in the literature.

We fit the spectrum of source~A with a power-law model with
an exponential cutoff,
$dN_{\rm ph}/dE_{\rm ph} = K E_{\rm ph}^{-\Gamma}\exp (-E_{\rm ph}/E_{\rm ph,cut} )$,  
typical for a pulsar \citep{Abdo2010Pul}.
The model fit yields $E_{\rm ph,cut} = 4.0\pm0.9\,{\rm GeV}$ 
and the spectral index of $\Gamma=1.9\pm0.1$.
In this model, the flux in energy range 0.2--300\,GeV
is $(7.5\pm0.9)\times10^{-11}\,\mbox{erg}\cm^{-2}\ps$, and the
corresponding luminosity is $(1.3\pm0.2)\times 10^{36}\du^{\,2}$ erg s$^{-1}$.
The significance of the exponential cutoff power law
(approximately described by
$\sqrt{TS_{cutoff}}\sigma = \sqrt{TS_{PL+cutoff} - TS_{PL}}\sigma$) 
is $\sim6\sigma$. 
The spectral shape of source~A is similar to
those of the detected \gray\ pulsars \citep{Abdo2013Pul}, 
which usually show flat spectra below 1~GeV and
exponential cutoffs in the energy range $\sim0.4$--$6\,$GeV.
If this source is a ``kicked" pulsar moving from the SNR center,
then the best-fit position, $0.^{\circ}1$ away, would imply a
projected traverse velocity
of 180--$4900\du\km\ps$ if the remnant's age estimate 4-110~kyr
\citep{gaoyuan2015} is adopted.
(The closer the position within the $3\sigma$ circle is to the SNR center, 
the lower the velocity would be.)
The upper limit of the velocity seems very high,
but there is also acuumulating evidence for high pulsar
velocities, even exceeding $4\times10^3\km\ps$
(e.g., PSR~B2011+38 and PSR~B1718-35, \citealt{zou2005}).
On the other hand, if it is an associated pulsar at $\sim12$\,kpc,
then its \gray\ luminosity of the order of $10^{36}\,\mbox{erg}\ps$
(Section \ref{subsec:sa})
would be among the highest among the (radio loud) \gray\ pulsars
\citep{Abdo2013Pul},
which seems difficult to accept in view of the no detection of
any radio pulsar here.

\subsection{Emission from Particles Accelerated by Kes~41?}
\begin{figure}
\centering
\includegraphics[width=0.85\textwidth]{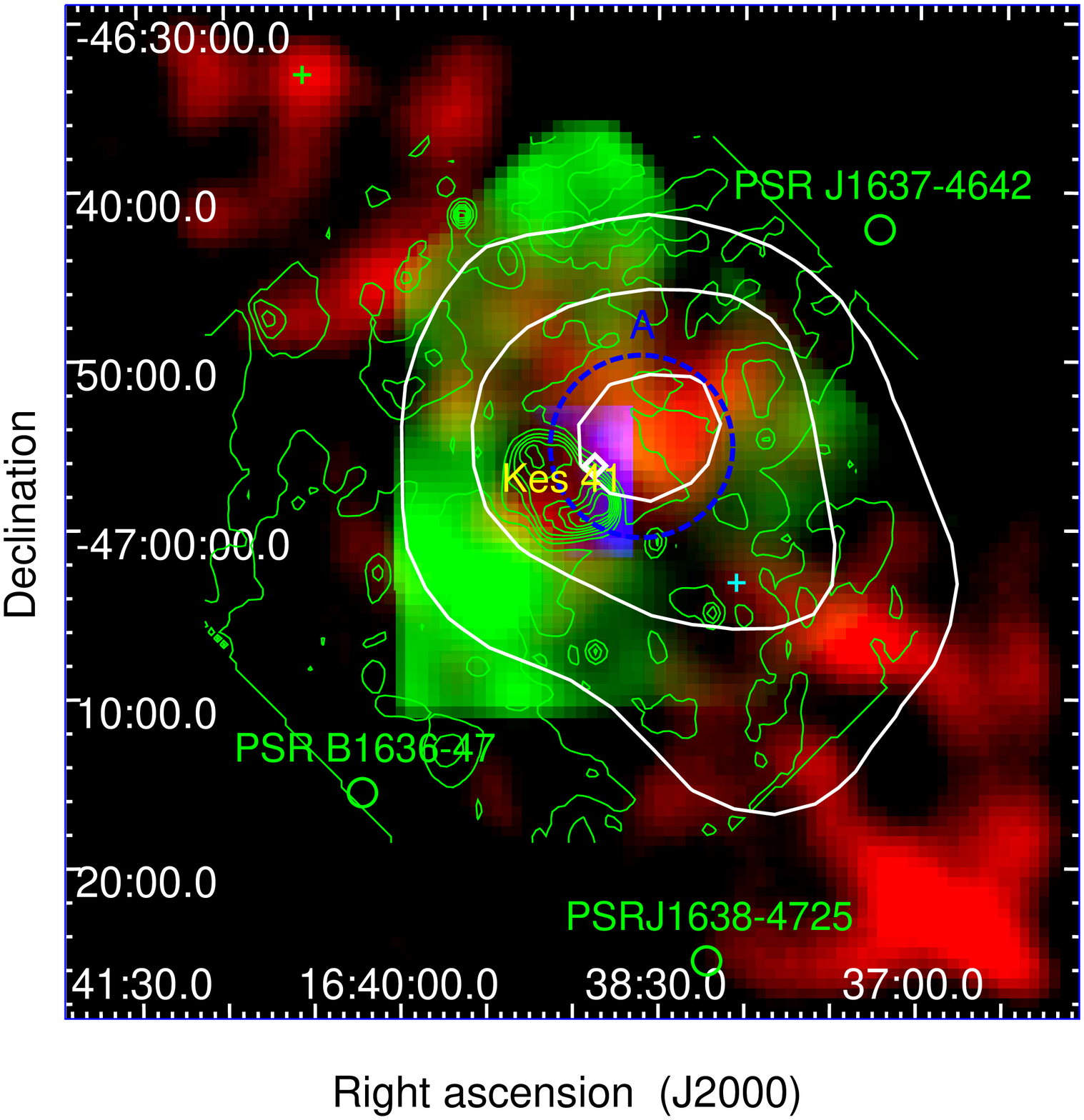}
\caption{Tri-color image of Kes~41 in multiwavelengths.
Red: \fermi-LAT 2--300GeV counts map centered at SNR Kes~41,
smoothed  with a Gaussian of width 0\fdg6
(per pixel bin representing 0\fdg01).
Blue: \twCO~(\Jotz) integrated emission ($\VLSR=-70$ to $-40\km\ps$)
with a field of view of $11'\times10'$.
Green: \HI\ line emission from SGPS integrated map ($\VLSR=-55$ to $-50\km\ps$).
The green contours and the green and cyan crosses
 are the same as in Fig.\ref{fig:tsmap}. 
The white curves show the TS~$=100, 144$ and 196 contours
(which correspond to significance 10$\sigma$, 12$\sigma$ and 14$\sigma$,
respectively).
The white diamond indicates the location of the $1720\MHz$ OH maser 
\citep{Koralesky1998}
and the green circles label the positions of known pulsars.
The dashed blue circle indicates the 3$\sigma$ error circle
of the best-fit position of source~A.
}
\label{fig:rgbmap}
\end{figure}

The $3\sigma$ error circle is on the northwestern boundary of
SNR~Kes~41 and essentially consistent with the
shock-MC interaction region.
Actually, it covers not only the 1720MHz OH maser but also the northwestern
molecular gas at a systemic velocity of $\VLSR\sim-50\km\ps$ 
that surrounds the remnant (\citealt{gaoyuan2015}; see Fig.~\ref{fig:rgbmap}).
It is very possible that the \gray\ emission arises from the
relativistic particles accelerated by the SNR shock waves.
We need to confront the leptonic and hardronic mechanisms
with the obtained \gray\ data.

\subsubsection{Leptonic Scenario}
First, we consider the scenario in which
the \gray\ emission comes from the inverse Compton scattering
off the relativistic electrons accelerated by the SNR shock.
The emissivity of the bremsstrahlung process is compatible with 
that of the {\em p-p} process if the number ratio of electrons to protons 
at a given energy, $K_{ep}$, is of order $\sim 0.1$ \citep{Gaisser1998}. 
Nevertheless, the values of $K_{ep}$ observed at Earth \citep{Yuan2012LB} 
and predicted by the diffusive shock acceleration theory \citep{Bell1978A} 
are both of the order of $\sim 0.01$. 
Therefore, the bremsstrahlung gamma-ray is usually insignificant.

We fit a power-law electron spectrum with a cutoff,
$dN_{\rm e}/dE_{\rm e} \propto E_{\rm e}^{-\alpha_{e}}\,
 {\rm exp}(-E_{\rm e}/E_{\rm e,cut})$,
to the spectral data and only consider the cosmic microwave
background as the seed photons
(referred to as {\em Case~A\/}).
As can be seen in Fig.~\ref{fig:sed} (blue dotted line), the fitting effect is less satisfactory.
We obtain $\alpha_{e}\approx2.0$ and $E_{\rm e,cut}\approx400$ GeV
(also see Table~\ref{tab:cases}).
The normalization is given by the total energy deposited in electrons
with energy above 1 GeV,
$W_{\rm e}(>1{\rm GeV}) \sim 1.3\times 10^{51}$ erg.
This electron energy budget is unreasonably high as the
order of the canonical SN explosion energy.

\subsubsection{Hadronic Scenario}
\begin{figure}
\centering
\includegraphics[width=0.85\textwidth]{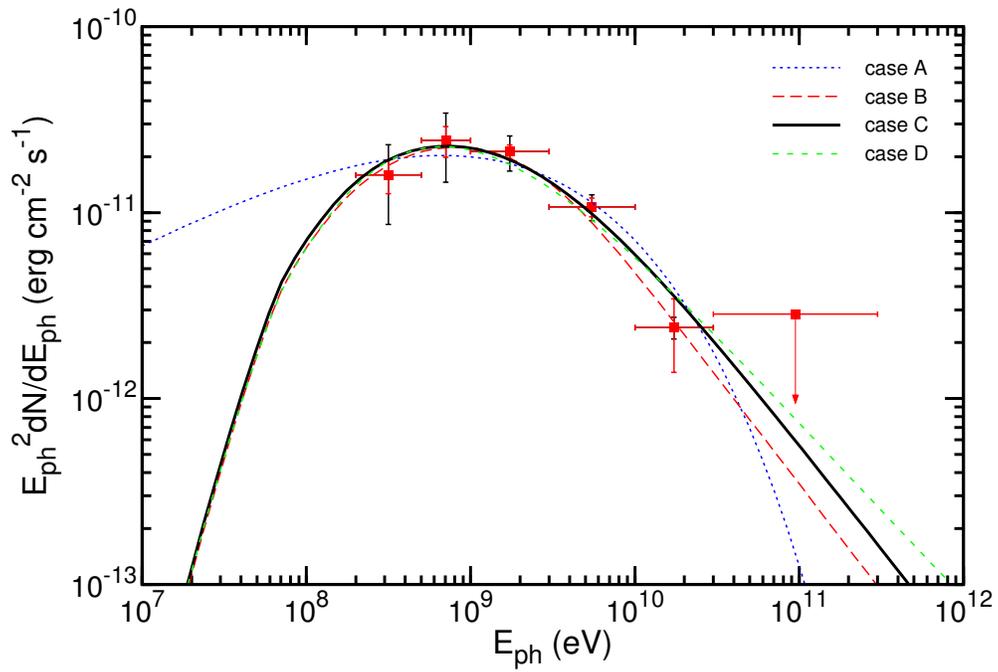}
\caption{\fermi\ \gray\ spectral energy distribution
 of source~A fit with various models (see text).
Systematic errors (see Section \ref{subsec:sa})
are indicated by black bars and the statistical errors 
are indicated by red bars.}
\label{fig:sed}
\end{figure}

Next, we  consider the scenario in which  the \gray\ emission is produced by the
collision of the shock accelerated protons with dense molecular gas.
For the case (referred to as {\em Case B}) in which the protons
collide with the dense target molecular gas (with average number density $\nt$),
we assume for the protons a broken power-law distribution,
$dN_{\rm p}/dE_{\rm p} \propto
E_{\rm p}^{-\alpha_{\rm p}}(1+(E_{\rm p}/E_{\rm b})^{2})^{-\Delta\alpha_{\rm p}/2}$,
to fit the spectral data (see Fig.~\ref{fig:sed} (red dashed line) and Table~\ref{tab:cases}).
We thus obtain
$\alpha_{\rm p} = 2.0$, $\Delta\alpha_{\rm p} = 1.2$, and a break energy of $E_{\rm b}=18$ GeV.
The total energy deposited in the protons with energy above 1~GeV is
$W_{\rm p}(> 1\,{\rm GeV})\sim 0.7\times10^{50}E_{51}(\nt/100\cm^{-3})^{-1}$ erg,
where $E_{51}=E_{\rm SN}/10^{51}\,\mbox{erg}$ is the dimensionless
SN explosion energy.
SNR~Kes~41 has been found to be surrounded by molecular gas
of density with $n_{\rm H_2}\sim140$--$500\cm^{-3}$ in the northwest
and HI gas of density with $n(\mbox{HI})\sim40\cm^{-3}$ in the southeast 
(also see Fig.~\ref{fig:rgbmap}).
If the mean target density $\nt$ is approximately of the order of $100\cm^{-3}$,
then $W_{\rm p}(> 1\,{\rm GeV})\sim1\times10^{50}E_{51}\,$erg,
namely, the fraction, $\eta$, of the SN explosion energy converted to protons
is of the typical order of 0.1.
While in this scenario the hadronic \gray{s} are emitted
at the SNR shock, it is noteworthy that the centroid of the $3\sigma$ circle
of source~A appears to be outside the northwestern boundary of the SNR.

The hadronic emission can alternatively be considered as originating from
the adjacent MCs that are ``illuminated" by the diffusive
relativistic protons escaping from the SNR shock front.
In the finite volume of a nearby cloud,
the protons' energy distribution can be obtained by calculating
the diffusive escaping protons accumulatively throughout
the history of the SNR expansion ({\em Case~C\/}).
For such a calculation, in the following,
we refer to \citet{lichen2012} and the references therein
for details of the model.

In the model calculation, we assume a converted CR proton energy fraction of
$\eta=0.1$ and an SN explosion energy of $E_{\rm SN}=10^{51}$\,erg.
The SNR radius in the southeast-northwest
orientation is adopted as $R_s\approx11\,$pc.
The \gray{s} are assumed to arise from an MC, of thickness $\Delta R_c$,
which is in contact with the shock surface;
therefore, the MC center is at $R_c=R_s+\Delta R_c/2$ away from the SNR center.
According to \citet{gaoyuan2015},
the SNR evolves in a cavity and may have been drastically decelerated
and entered the radiative phase as soon as the blast wave
encountered the cavity wall,
after a Sedov evolution lifetime \citep{Sedov1959} of
$t_{\rm enc} = 4\times 10^{3} (n_{\rm H}/0.3\cm^{-3})^{1/2}
               E_{51}^{-1/2}(R_s/11\,{\rm pc})^{5/2}$~years.
We assume that the particle acceleration process is
not significant after this time.
Therefore, the average distribution of the cumulative escaping protons
in the volume of the MC at the remnant age $t_{\rm age}$ is rewritten as
\begin{eqnarray}
F_{\rm ave}(E_{\rm p},t_{\rm age})
   & = & \int^{R_c+\Delta R_c/2}_{R_c-\Delta R_c/2}r^2dr
      \int^{t_{\rm enc}}_{0}\int^{2\pi}_{0}\int^{\pi}_{0}
       f(E_{\rm p}, R_{\rm bet}(R_{\rm c},t_{i},\theta,\phi),t_{\rm dif})
      R_{\rm s}^{2}(t_{i}){\rm sin}\theta \,d\theta \,d\phi \,dt_{i}\nonumber\\
   & & \Big/ \int^{R_c+\Delta R_c/2}_{R_c-\Delta R_c/2}r^2dr, 
\end{eqnarray}
where
$t_{i}$ is the time at which a proton escapes from the SNR shock,
$t_{\rm dif}=t_{\rm age}-t_{i}$ is the diffusion time after escape,
$R_{\rm bet}$ is the distance between the escape point on the shock surface
and a given point in the cloud (with position angles ($\theta$, $\phi$)),
and $f(E_{\rm p}, R_{\rm bet}(R_{\rm c},t_{i},\theta,\phi),t_{\rm dif})$
is the distribution function at a given point of the protons
that escape from the unit area at an arbitrary escape point.
Considering the remnant's age range $\sim$4--100\,kyr estimated from
the ionization timescale of the X-ray emitting gas \citep{gaoyuan2015},
we calculate the model with three age numbers, 4, 10, and 100\,kyr.
This model can fit the spectral points as well,
as exemplified by the solid line for $t_{\rm age}=10$\,kyr
in Figure~\ref{fig:sed}.
The model parameters are listed in Table~\ref{tab:cases}.
The photon index $\alpha_{\rm p}=2.4$,
the energy-dependent index of the diffusion coefficient $\delta=0.7$,
and the correction factor of slow diffusion around the SNR $\chi\sim0.01$--0.1
are in normal ranges.
The $\Delta R_c$ value $\sim5$--13\,pc ($\sim0\fdg02$--$0\fdg06$)
implies that the MC
involed in the {\em p-p} hadronic interaction is essentially
within the $3\sigma$ circle of source~A.
Note that such a source size is much smaller than the PSF
size $0\fdg5$ of the \fermi-LAT at energies above 2\,GeV,
 consistent with the above judgement of
a point-like source.
The ``illuminated" MC mass $M_{\rm cl}$, around $10^5M_{\odot}$,
seems reasonable as compared with the mass of the molecular gas
``reservoir" in the northwest, which is  no less than
$\sim$ a few times $10^5M_{\odot}$,
which is estimated from a limited field of view of the CO observation
\citep{gaoyuan2015}.

However, the cavity wall may send a reflected shock backward
when the blast wave collides with it \citep{gaoyuan2015}.
If the reflected shock can still effectively accelerate particles
after the forward shock becomes radiative,
then the situation would be more complicated than the above cases.
For simplicity, we approximate this case as a continuous
proton injection from the SNR center \citep{Aharonian1996AA}
({\em Case~D\/}).
In this case, the MC is regarded as a point at $R_c$ from the SNR center
and the same energy conversion fraction $\eta=0.1$ is adopted.
We follow the algorithm described in \citet{Aharonian1996AA}
and fit the spectral data, as exemplified by the green dashed line
for $t_{\rm age}=10$\,kyr and $R_c=20$\,pc
in Figure~\ref{fig:sed}.
These model results are generally similar to those of {\em Case~C\/},
with a slightly harder model spectrum at $\ga100$\,GeV.
For the three sets of parameters with $t_{\rm age}=10$ and 100\,kyr,
we again have $\alpha_{\rm p}=2.4$ and $\delta=0.7$.
The $\chi$ values are $\sim0.05$--0.5 in a normal range.
A higher mass of the ``illuminated" part of MC than {\em Case~C\/}
is required, but is still consistent with the MC mass estimate
from the CO observation in the order of magnitude.

The hadronic scenarios, both the interaction at the shock ({\em Case~B\/})
and the illumination by escaping protons ({\em Case~C/D\/}),
can generally explain the \gray\ properties of source~A.
The escape cases have harder model spectra at $\ga10$\,GeV
than the interaction-at-the-shock case.
Further TeV observations will likely be of help to distinguish
the two scenarios.

\section{COMPARISON WITH OTHER GeV-DETECTED SNRs IN MC ENVIRONMENTS}

We now present a brief discussion of Kes 41 within the context of other
Galactic SNRs that have been detected at $\gamma$-ray energies. An intriguing
trend has emerged in these studies where the Galactic SNRs that are known to
be interacting with dense clouds and that are detected at (very) high energies
also appear to exhibit contrasting morphologies in the X-ray and the radio.
Specifically, these sources exhibit the shell-like radio morphologies that
are characteristic of SNRs coupled with a center-filled X-ray morphology that
is thermal in origin,
and therefore belong to the class of thermal composite or mixed-morphology
SNRs (also see Section \ref{sec:intro}).
Actually, about half of the 36-37 known thermal composites
have been found to be interacting with adjacent MCs
\citep[see Table~4 in][]{gaoyuan2015}.
While the origin of these contrasting morphologies remains uncertain,
it appears that the interaction between the SNRs and the dense clouds plays a crucial
role. Proposed origins for these morphologies include the 
evaporation of shock-engulfed cloudlets,
thermal conduction within the interior hot gas,
and heating by the shock reflected from the wind-cavity wall;
the reader is referred to \citet{Chen08}
and references therein for a detailed review
of these proposed mechanisms.

We tabulate the thermal composite SNRs that have been detected at
$\gamma$-ray energies by \fermi-LAT in Table~\ref{tab:mmsnrs}.
So far, there are 13 SNRs (including Kes~41) of this class that
have associated GeV \gray\ emission, and an additional six of them 
possibly have associated GeV \gray\ emission,
as listed in Table~\ref{tab:mmsnrs}.
We can see that most of the GeV-detected thermal composites are in physical
interaction with MCs.

In Table~\ref{tab:mmsnrs}, we collect the photon indices in the GeV band
and adopt the luminosities in, or convert them to, the 1--100 GeV
energy range for ease of comparison.
The $\sim$GeV spectra of these SNRs are soft,
with power-law photon indices of $\Gamma\ge2.0$,
in distinct contrast with the hard spectra ($\Gamma\sim1.4$--1.8)
of the supposed leptonic process dominated \gray\ SNRs,
e.g., 
RX~J0852.0--4622 \citep{Tanaka2011} and
RCW~86\citep{Yuan2014}.
Except for HB~21 and Kes~27, the 1--100~GeV luminosities
of the 13 identified GeV \gray\
sources are on the order of a few times $10^{35}\
{\rm erg\, s^{-1}}$, which are significantly higher than those of
the leptonic process dominated SNRs (e.g., 
$< 10^{34}\ {\rm erg\, s^{-1}}$ for RX~J0852.0--4622, \citealt{Tanaka2011};
and RCW~86, \citealt{Yuan2014}).
For exceptional cases of HB~21 and Kes~27, the low luminosities may be
due to proton collisions with only a very small amount of
dense clouds \citep[e.g.,][]{Pivato2013, Xing2015Kes27}.
We note that where detailed modeling has been applied to
the $\gamma$-ray spectra of these sources, hadronic models have
generally proved to give better fits to the data than leptonic models
(except for the uncertain cases of Kes 17 and HB 9).

These past $\gamma$-ray observations of thermal composite SNRs--including 
the observation of Kes 41 that is presented in this paper -- have thus produced 
insights into how SNRs interact with MC and how SNRs accelerate CR
particles. Additional $\gamma$-ray observations of thermal composites
are necessary and timely to explore the relation between emission at these high energies
and the origin of the contrasting morphologies that characterize SNRs of this type.

\section{Summary}
 
We perform an analysis of the \gray\ emission in a
$14^{\circ}\times14^{\circ}$ region centered on the thermal composite
SNR~Kes~41, using 5.6 years of \fermi-LAT observation data.
We find a point-like source to the northwest of the SNR
with a significance of $24\sigma$ in 0.2--300\,GeV.
Neither significant long-term variability nor periodicity 
is detected from the timing analysis of source A in the same energy range.
The $3\sigma$ error circle, $0\fdg09$ in radius,
covers the 1720MHz OH maser
and is essentially consistent with
the location of the $\VLSR\sim-50\km\ps$ MC with which
the SNR interacts.
The source emission can be described by a
power-law spectrum with an exponential cutoff
with a photon index of $1.9\pm0.1$ and
a cutoff energy of $4.0\pm0.9\ {\rm GeV}$.
The corresponding 0.2--300\,GeV flux is $(7.5\pm0.9)\times10^{-11}
\,\mbox{erg}\cm^{-2}\ps$, and the luminosity is
$\sim1.3\times 10^{36}\,\mbox{erg}\ps$ at a distance of 12\,kpc.
Although the spectrum is similar to those of pulsars,
there is no radio pulsar in the $3\sigma$ circle
responsible for the high luminosity.
While the power-law electron spectrum with a cutoff
for inverse Compton scattering would lead to a difficulty in
the electron energy budget,
the emission can be naturally explained by the
hadronic interaction between the relativistic protons
accelerated by the shock of SNR~Kes~41 and the adjacent northwestern MC.
By comparison with the hadronic interaction at the shock,
which appears off the best-fit position of the source,
illumination of the adjacent MC by the protons escaping from
the shock front seems more consistent with observations.
A list of Galactic thermal composite SNRs detected
at GeV \gray\ energies by \fermi-LAT is presented in this paper.

\bigskip B.L. is grateful to Xia Fang, Ning-Xiao Zhang, and Zheng-Gao Xiong 
for the help about {\it Fermi} data analysis. We thank the support of
NSFC grants 11233001 and 11403075. This work 
has also benefited from 973 Program grant 2015CB857100, 
grant 20120091110048 from the Educational Ministry of China,
and the grants from the 985 Project of NJU and the Advanced
Discipline Construction Project of Jiangsu Province.
This research has made use of the SIMBAD database,
operated at CDS, Strasbourg, France.

\bibliographystyle{apj}
\bibliography{cite}

\begin{deluxetable}{cccc}
\tablecaption{\fermi\ LAT Flux Measurements of Source A in the Kes~41 Region}
\tablewidth{0pt}
\startdata
\hline
\hline
$E_{\rm ph}$ (energy band) & 
  $E_{\rm ph}^2dN(E_{\rm ph})/dE_{\rm ph}$\tablenotemark{a} & TS value \\
 (GeV)         & (10$^{-12}$ erg cm$^{-2}$ s$^{-1}$) &  \\
\hline
0.32 (0.20--0.50) &    15.9$\pm$3.3$\pm$7.3    & 39  \\
0.71 (0.50--1.00) &    24.4$\pm$4.6$\pm$9.9    & 113\\
1.73 (1.00--3.00) &   21.3$\pm$1.8$\pm$4.6    & 224\\
5.48 (3.00--10.0) &   10.7$\pm$1.2$\pm$1.7    & 105\\
17.3 (10.0--30.0) &   2.4$\pm$1.0$\pm$0.3  & 9 \\
94.9 (30.0--300)  &   $\le2.9$\tablenotemark{b} & 2\\
\enddata
\tablenotetext{a}{The first column of errors lists statistical errors and
the second lists systematic errors.}
\tablenotetext{b}{The 95$\%$ upper limit.}
\label{tab:flux}
\end{deluxetable}

\begin{deluxetable}{c|ccccccc}
\tablecaption{Model Parameters for the Emissions of SNR Accelerated Particles}
\tablewidth{0pt}
\startdata
\hline
\hline
  & & $\alpha_{\rm e}$ & & $E_{\rm e,cut}$ & $W_{\rm e}(>1\mbox{GeV})$ \\
  & &        &          & (GeV)     &  ($10^{51}$ erg)\\
\hline
case A & & 2.0 & & 400 & 1.3 \\
\hline
\hline
  & & $\alpha_{\rm p}$ & $\Delta\alpha_{\rm p}$ & $E_{\rm b}$ & $\nt E_{51}^{-1}W_{\rm p}(>1\mbox{GeV})$ \\
  & &       &         & (GeV) & ($10^{51}$ erg$\cm^{-3}$)\\
\hline
case B & & 2.0 & 1.2 & 18 & 7 \\
\hline
\hline
 &$t_{\rm age} $& $\alpha_{\rm p}$ &$\delta$ & $\chi$ & $\Delta R_c$ & $M_{\rm cl}$ \\
 & (kyr) & & & &(pc) & $(10^{4}\, M_{\odot})$ \\
\hline
case C &4   & 2.4 & 0.7 & 0.07  & 5  & 4.5 \\
       &10  & 2.4 & 0.7 & 0.03  & 10 & 11  \\
       &100 & 2.4 & 0.7 & 0.004 & 13 & 18  \\
\hline
\hline
 &$t_{\rm age} $& $\alpha_{\rm p}$ &$\delta$ & $\chi$ & $R_c$ & $M_{\rm cl}$ \\
 & (kyr) & & & &(pc) & $(10^{4}\, M_{\odot})$ \\
\hline
case D &10  & 2.4 & 0.7 & 0.25  & 15 & 18\\
       &10  & 2.4 & 0.7 & 0.45  & 20 & 40\\
       &100 & 2.4 & 0.7 & 0.05  & 20 & 40
\enddata
\label{tab:cases}
\end{deluxetable}

\begin{deluxetable}{cccccc}
\tabletypesize{\scriptsize}
\tablecaption{Parameters of the \gray\ Emission of the Galactic Thermal
Composite SNRs Obtained from \fermi--LAT Observation}
\tablewidth{0pt}
\tablehead{
\colhead{Source} & \colhead{Distance} & \colhead{$\Gamma$}& \colhead{$L_{1-100\, {\rm GeV}}$} & \colhead{MC interaction$^{\rm a}$} & \colhead{References} \\
 & \colhead{(kpc)} & & ($10^{35}\ {\rm erg\,s^{-1}}$) &
}
\startdata
G6.4--0.1(W28)         & 2.0 & $2.74\pm0.06$\tablenotemark{b} & 1.0  & Y & 1,2 \\
G31.9+0.0(3C 391)      & 7.2 & $2.50\pm0.04$\tablenotemark{b} & 4.0  & Y & 3,4   \\
G34.7--0.4(W44)        & 2.8 & $3.02\pm0.10$\tablenotemark{b} & 2.7  & Y & 5,6 \\
G43.3--0.2(W49B)       & 8   & $2.29\pm0.02$\tablenotemark{c} & 8.0 & Y  & 7,8\\
G49.2--0.7(W51C)       & 6   & $2.5 \pm0.1 $\tablenotemark{b} & 4.4 & Y   &9,10\\
G89.0+4.7(HB 21)       & 1.7 & $2.33\pm0.03$\tablenotemark{c} & 0.13 & Y  &11,12\\
G189.1+3.0(IC 443)     & 1.5 & $2.61\pm0.04$\tablenotemark{b} & 1.0  & Y  &5,13\\
G304.6+0.1(Kes 17)     & 9.7 & $2.0 \pm0.3 $\tablenotemark{c} & 12   & Y  &14,15\\
G327.4+0.4(Kes 27)     & 4.3 & $2.5 \pm 0.1$\tablenotemark{c} & 0.24 &   &16,17\\
G337.8--0.1(Kes 41)    & 12  & $2.38\pm0.03$\tablenotemark{c} & 7.7  & Y  &18\\
G348.5+0.1(CTB~37A)    & 11.3& $2.19\pm0.07$\tablenotemark{c} & 7.8  & Y &19,20,21\\
G357.7--0.1(MSH 17-39) & 12  & $2.5 \pm0.3 $\tablenotemark{c} & 5.8  & Y &22,23\\
G359.1--0.5            & 7.6 & $2.60\pm0.05$\tablenotemark{c} & 4.0  & Y &24,25\\ 
\hline
G0.0+0.0(Sgr A East) (?)$^{\rm d}$   & 8.0 & $2.32\pm0.03$\tablenotemark{c} & 8.7 & Y &26,23\\
G132.7+1.3(HB 3) (?)$^{\rm d}$      & 2.2 & $2.30\pm0.11$\tablenotemark{c} & 0.04 &Y? &27,23\\
G156.2+5.7 (?)$^{\rm d}$            & 3   & $2.35\pm0.09 $\tablenotemark{c} & 0.26 & & 28,23\\
G290.1--0.8(MSH 11-61A) (?)$^{\rm d}$& 7   & $\sim2.28       $\tablenotemark{c} & 1.5 & ? &22,23\\
\hline
G160.9+2.6(HB 9) (?)$^{\rm e}$        & 1.0 & $2.30\pm0.05$\tablenotemark{c} & 0.013 & ?&29\\
G166.0+4.3 (?)$^{\rm e}$            & 4.5 & $2.27 \pm0.1 $\tablenotemark{c} & 0.11 & ? &30,31\\
\enddata
\tablecomments{(1) \citealt{Frail2011};
(2) \citealt{Abdo2010W28}; 
(3) \citealt{Radhakrishnan19723};
(4) \citealt{Ergin2014T}; 
(5) \citealt{Seta1998H};
(6) \citealt{Abdo2010W44SCi};
(7) \citealt{Brogan2001}; 
(8) \citealt{Abdo2010W49B}; 
(9) \citealt{Koo2005L}
(10) \citealt{Abdo2009W51C}; 
(11) \citealt{Byun2006};
(12) \citealt{Pivato2013}
(13) \citealt{Abdo2010IC443};
(14) \citealt{Combi2010AS};
(15)\citealt{Gelfand2013CS};
(16)\citealt{McClure2001};
(17)\citealt{Xing2015Kes27};
(18)\citealt{gaoyuan2015};
(19)\citealt{Reynoso2000};
(20)\citealt{Castro2010};
(21)\citealt{CTB37A2013fermi};
(22)\citealt{Rosado1996};
(23)\citealt{Acero2015};
(24)\citealt{Uchida1992};
(25)\citealt{HuiCY2011};
(26)\citealt{Reid1993MJ};
(27)\citealt{Routledge1991};
(28)\citealt{Reich1992};
(29)\citealt{Araya2014};
(30)\citealt{Landecker1989};
(31)\citealt{Araya2013};
}
\tablenotetext{a}{Adopted from \citealt{Jiang2010CW} SNR--MC association table.}
\tablenotetext{b}{The photon index above the break energy for broken power-law spectrum.}
\tablenotetext{c}{The photon index of single power-law spectrum.}
\tablenotetext{d}{Question mark: the association of the detected \gray\ emission with the SNR is uncertain.}
\tablenotetext{e}{Question mark: not listed in the latest 3FGL catalogue \citep{Acero2015}.}
\label{tab:mmsnrs}
\end{deluxetable}

\end{document}